\documentclass{iaus}
\usepackage{graphicx,natbib,subfigure}
\DeclareGraphicsExtensions{.eps}
\title[Hot gas and dark matter in galaxy clusters] 
{Investigating the Relationship Between the Hot Gas and the Dark Matter Components of Galaxy Clusters.}

\author[Leila C.  Powell \textit{et al.}]   
{Leila C. Powell$^1$, 
 Scott T. Kay$^2$, Arif Babul $^3$\break \and Andisheh Mahdavi$^3$}

\affiliation{$^1$Oxford Astrophysics, Denys Wilkinson Building, Keble Road, Oxford, OX1 3RH, UK \break email: lcp@astro.ox.ac.uk\\[\affilskip]
$^2$Jodrell Bank Observatory, University of Manchester, Macclesfield, Cheshire,  SK11 9DL, UK

$^3$University of Victoria, Department of Physics and Astronomy, Victoria, V8P 5C2, Canada}

\pubyear{2007}
\volume{244}  
\pagerange{451--452}
\date{?? and in revised form ??}
\jname{Dark Galaxies and Lost Baryons}
\editors{J. I. Davies \& M. J. Disney, eds.}
\begin{document}

\maketitle

\begin{abstract}
Various differences in galaxy cluster properties derived from X-ray and weak lensing observations have been highlighted in the literature. One such difference is the observation of mass concentrations in lensing maps which have no X-ray counterparts (e.g. \citealt{jee:apj}). We investigate this issue by identifying substructures in maps of projected total mass (analogous to weak lensing mass reconstructions) and maps of projected X-ray surface brightness for three simulated clusters. We then compare the 2D mass substructures with both 3D subhalo data and the 2D X-ray substructures.
Here we present preliminary results from the first comparison, where we have assessed the impact of projecting the data on subhalo identification.

\end{abstract}

\firstsection 

\section{Introduction}
Six clusters from a $\Lambda$CDM cosmological simulation (performed by the Virgo Consortium\footnote{http://www.virgo.dur.ac.uk}), are resimulated, using GADGET-2 \citep{gadget2}, with 10$^5$ particles (low resolution) and a non-radiative gas physics model.
In order that we can later assess the impact of dynamical state, three are selected for their varied mass histories and are also resimulated with $\sim$ 1x10$^6$ particles (high resolution). We examine the clusters within r$_{500}$ as this is an approximate upper limit on the extent of X-ray observations.

\section{Identifying substructures in 2D and 3D}
The maps are smoothed with a Gaussian filter of FWHM= 60 kpc h$^{-1}$ to remove structures smaller than the simulation's force resolution and reduce random noise.
The substructures in the maps are then enhanced using an unsharpmasking technique to remove the cluster background.
 The enhanced map is searched for substructures with greater than 3$\usigma$ significance, that are then characterised by ellipses.
This technique is effective for relaxed and disturbed clusters as it does not rely on circular symmetry.

Positions and masses of self-bound 3D subhaloes are obtained using a version of SUBFIND \citep{subfind:orig} and these are then cross-referenced with the catalogues of 2D mass substructures.
A subhalo is considered a match to a source in the map if its centre lies within the ellipse that characterises the source (see figure \ref{fig2} for matching success).
If more than one subhalo satisfies this criteria, the source in the map is flagged as a Ômultiple matchÕ and is assigned the combined mass of all the subhaloes. 

\section{The effect of projection on substructure identification}\label{sec:projection}
\begin{figure}[h]
  \hfill
  \begin{minipage}[t]{.45\textwidth}
    \begin{center}  
      \includegraphics[width=6cm]{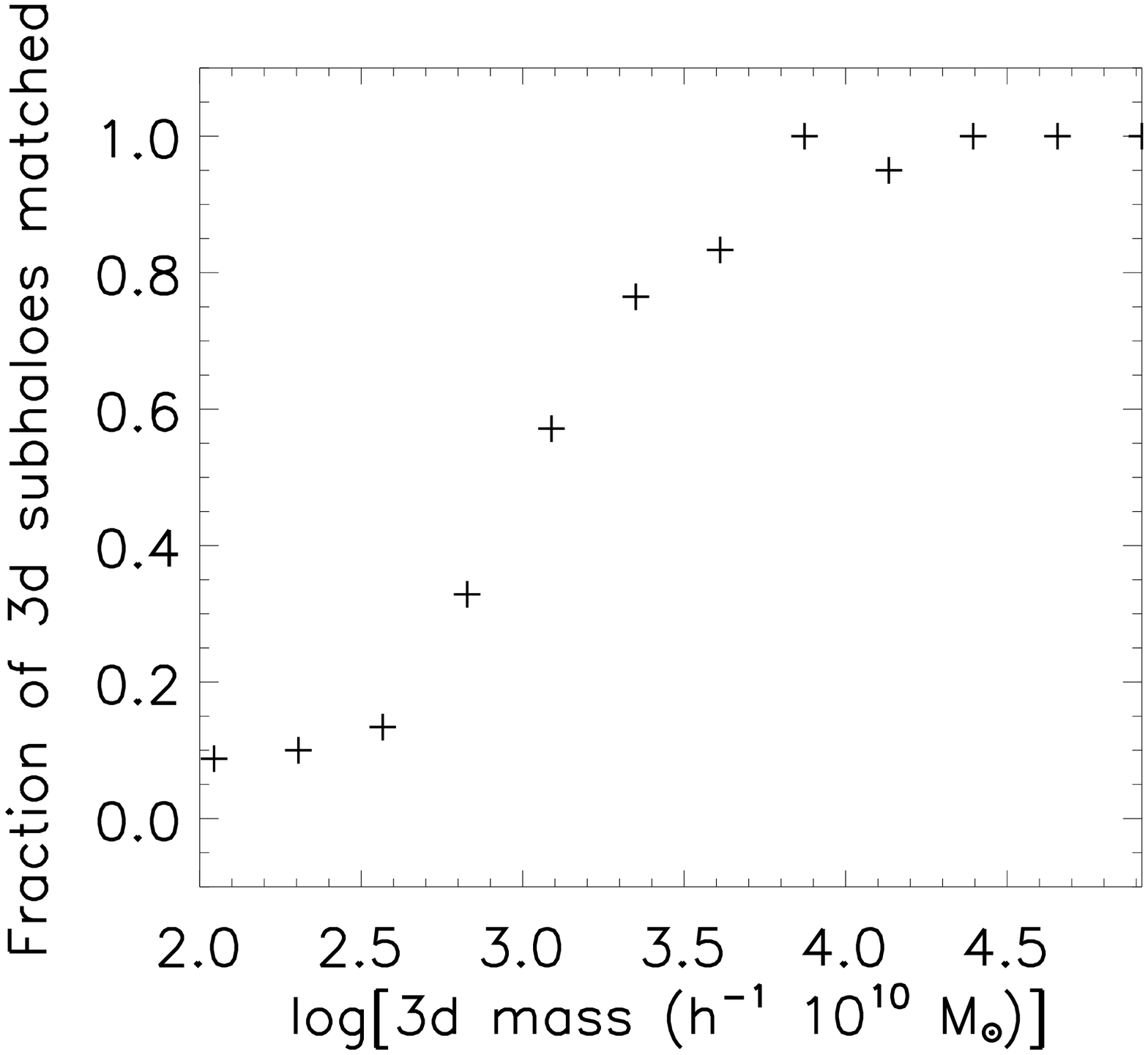}
      \caption{Dependence of matching success of 3D subhaloes to 2D substructures in the projected mass map on subhalo mass (all clusters combined).}
      \label{fig2}
    \end{center}
  \end{minipage}
  \hfill
  \begin{minipage}[t]{.45\textwidth}
    \begin{center}  
      \includegraphics[width=6cm]{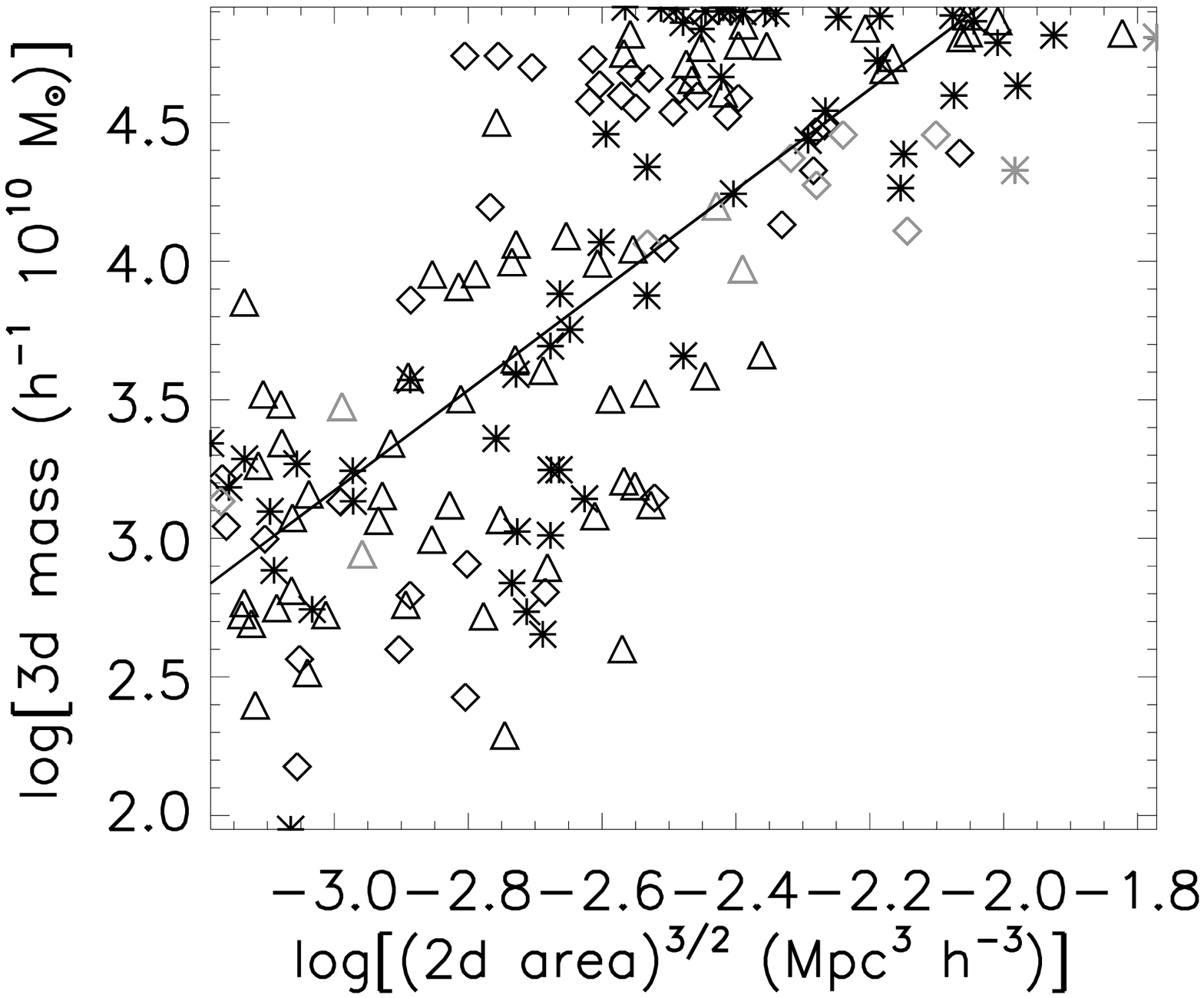}

      \caption{Correlation between area of 2D substructure in map and associated 3D subhalo mass for cluster 1 (asterix) , cluster 2 (diamonds) and cluster 3 (triangles). 2d substructures which are a multiple match (with 100:1 mass limit) are shown in grey.}
      \label{fig1}
    \end{center}
  \end{minipage}
 
  \hfill
\end{figure}

We find a clear correlation between the area of the substructure in the enhanced projected mass map and the subhalo mass(es) attributed to it after matching with the 3D data (figure \ref{fig1}). 
We apply a mass limit to the multiple matching technique, such that further subhaloes can only be matched to a source if they have a given mass ratio with the most massive matched subhalo. A significant number of multiple matches are found only if the limiting mass ratio is reduced to 100:1. This indicates that genuine projection effects are not an issue in our sample.

\section{Future work}\label{sec:greenfun}
We will investigate the impact of confusion, by varying the FWHM of the Gaussian filter and examining the effect on the probability of multiple matches.
By comparing the 2D mass substructures to the 2D X-ray substructures, we can establish how the two components differ and how this depends on 3D subhalo mass, dynamical state, position in the cluster and redshift.
This comparison will be extended to the high resolution simulations and to simulations including feedback and cooling (Powell \textit{et al.}, in preparation).

\begin{acknowledgments}
LCP is supported by an STFC studentship.
\end{acknowledgments}


\begin{thebibliography}{}
\bibitem[ Jee, White, Ford \etal\ (2005)]{jee:apj}
     {Jee, M. J., White, R. L., Ford, H. C., Blakeslee, J. P., Illingworth, G. D., Coe, D. A., Tran, K.-V. H.} 2005,
     \textit{ApJ} 634, 813-832
      \bibitem[Springel(2005)]{gadget2}
     {Springel, V.} 2005,
     \textit{MNRAS} 364, 1105-1134 
\bibitem[Springel, White, Tormen \etal\ (2001)]{subfind:orig}
     {Springel V., White, S. D. M., Tormen, G., Kauffmann, G.} 2001,
     \textit{MNRAS} 328, 726-750

 \end{thebibliography}
\end{document}